# Nonlatching Superconducting Nanowire Single-Photon Detection with Quasi-Constant-Voltage Bias

Deng-Kuan Liu[1,2], Si-Jing Chen[1,2], Li-Xing You[1,*], Yong-Liang Wang[1], Shigehito Miki[3], Zhen Wang[1,3], Xiao-Ming Xie[1], and Mian-Heng Jiang[1]

[1]*State Key Laboratory of Functional Materials for Informatics, Shanghai Institute of Microsystem and Information Technology, Chinese Academy of Sciences, 865 Changning Rd., Shanghai 200050, China*

[2]*Graduate University of the Chinese Academy of Sciences, Beijing 100049, PR China*

[3]*Kobe Advanced Research Center, National Institute of Information and Communications Technology, 588-2, Iwaoka, Nishi-ku, Kobe, Hyogo 651-2492, Japan*

**Abstract:** Latching is a serious issue in superconducting nanowire single-photon detector (SNSPD) technology. By extensively studying the electrical transportation characteristics of SNSPD with different bias schemes, we conclude that latching is a result of the improper bias to SNSPD. With the quasi-constant-voltage bias scheme, the intrinsic nonlatching nature of SNSPD is observed and discussed. The SNSPD working in the nonlatching bias shows a smaller jitter and a higher pulse amplitude than that in the previous anti-latching method. The quantum efficiency of SNSPD with the pulsed photon frequency up to 3 GHz is measured successfully, which further proves the nonlatching operation of SNSPD.

[*] E-mail address: lxyou@mail.sim.ac.cn

Over the past decade, the superconducting nanowire single-photon detection/detector (SNSPD) has attracted much attention in not only superconducting electronics but many other fields due to its great potential for applications such as quantum key distribution (QKD),[1-3] CMOS integrated circuits diagnostics,[4] and time-of-flight mass spectroscopy.[5] Compared with avalanche photodiodes (APDs) and photomultiplier tubes (PMTs), SNSPD has many merits in near-infrared wavelength such as a negligible dark count rate (<10 Hz), a small timing jitter (~50 ps), a high counting rate (>1 GHz), and a high quantum efficiency (>25%).[6,7] For practical application of SNSPD, there are a few issues that need to be solved or further studied, for example, the photon number resolution ability, the kinetic inductance, the timing jitter, and the latching behavior.

A quasi-constant-current (QCC) bias is widely employed in SNSPD electronics to provide a stable bias to SNSPD, which is often realized using an isolated voltage source in series with a resistor with a large resistance. The bias current is typically set close to its critical current $I_c$ (0.9-0.95$I_c$) to maintain a high quantum efficiency (QE) and a reasonable dark count rate ($R_{dc}$). When a photon hits the nanowire, a local resistive hotspot in the nanowire appears and produces a voltage pulse on SNSPD, which can be amplified by a room-temperature wideband amplifier and finally detected by a photon counter. In a very short period (tens of picoseconds), the hotspot's energy dissipates through the substrate and along the nanowire by phonon interaction. Then, the hotspot cools and the nanowire recovers to the superconducting state. The current in the nanowire decreases to a finite value quickly and then recovers to the original value slowly in the time scale determined by the kinetic inductance of SNSPD. SNSPD will be ready for the arrival of the next photons.

The above description shows an ideal continuous detection process without any disturbance. The practical operation of SNSPD suffers from an effect called latching,[8,9] which means an irreversible transition from the superconducting state to the resistive state under some disturbances. Intrinsic vortices in the nanowire, electrical noise, thermal fluctuation, and continuous photons with a short interval can be the trigger sources of latching. Indeed, the positive feedback energy provided by the bias circuit drives SNSPD into the resistive region. There are two disadvantages of the latching behavior. First, latching makes SNSPD very sensitive to the environmental noises that break the continuous operation of SNSPD and should be avoided in the practical SNSPD system. Second, once the photon repetition rate in the QKD experiment surpasses the maximal count rate of SNSPD, SNSPD cannot respond to the second continuous photon as long as the first photon response happens. Even worse, when random stray light acts like a transient photon train with high repetition rate, latch may also happen when the actual count rate is much less than the maximal count rate. This phenomenon seriously limits the working frequency of the practical QKD system using SNSPD. Besides, the latching effect seems to be more serious in the parallel nanowire detector. Suitable on-chip resistance or inductance was added to weaken the latching effect.[10-12]

In fact, as long as the positive feedback energy disappears, latching may not happen, though the trigger sources still exist. Driven by this idea, the *dc* current-voltage (I-V) characteristics of SNSPD are extensively studied under several different bias schemes, from a quasi-constant-voltage (QCV) bias to the popular QCC bias. We observed the nonlatching characteristics of SNSPD with the QCV bias, which is meaningful for not only the practical application of SNSPD but also understanding the detection mechanism.

The SNSPDs under study are fabricated with a 4-nm-thick NbN film deposited on an R-plane sapphire substrate. The active area of SNSPD is a nanowire of meander structure with a line width of 100 nm, which covers an area of $10 \times 10$ μm$^2$ with a filling factor of 0.5. The fiber-coupled SNSPDs are then installed in a multichannel Gifford-McMahon cryocooler with the working temperature of 2.7 K ± 20 mK. The superconducting transition temperature $T_c$ of these detectors is measured to be around 9.5 K. The SNSPD shows a typical quantum efficiency (QE) of 1% when the dark count rate $R_{dc}$ is 10 Hz for the wavelength of 1550 nm.

The electronics of SNSPD shown in Fig. 1 include a bias circuit and a readout circuit, which are connected to the detector through a bias-tee (bandwidth: 200 K-12 GHz) and a coaxial cable with the impedance $Z_0$=50 Ω. All the electronics work at room temperature except the SNSPD itself. The bias circuit is often composed of an isolated voltage source and a series resistor $R_s$ (≥10 KΩ), which forms a popular QCC bias to SNSPD via the *dc* arm of the bias-tee. The readout circuit includes a low-noise wideband amplifier and a counter/oscilloscope connected to SNSPD through the *rf* arm of the bias-tee. SNSPD is represented as an inductance in series with a resistor $R_n(t)$, whose value varies with time and depends on the current through SNSPD and the local temperature of the nanowire. To change the QCC bias to other modes including the QCV bias, we insert a parallel resistor $R_p$ between the *dc* arm of the bias-tee and the ground. When $R_p$ is far less than $R_s$ and the minimal resistance of the resistive SNSPD, the bias scheme works as a QCV bias to SNSPD.

To investigate the *dc* I-V characteristics of SNSPD, we measured the voltages of $R_s$ ($V_s$) and $R_p$ ($V_p$). Then, the current through SNSPD ($I_{SNSPD}$) and the voltage on SNSPD ($V_{SNSPD}$) can be calculated using the following equations:

$$I_{SNSPD} = \frac{V_s}{R_s} - \frac{V_p}{R_p}, \qquad (1)$$

$$V_{SNSPD} = V_p - I_{SNSPD} \cdot R_c. \qquad (2)$$

Indeed, $V_{SNSPD}$ is approximately equal to $V_p$ since $I_{SNSPD} \cdot R_c$ is around 0.1 mV, which is much less than $V_p$ in most cases. Here, we ignore the inductance in the bias-tee since it has little influence on the *dc* measurement.

By varying the resistance of $R_p$, we obtained I-V curves of SNSPD with different bias schemes. Four typical curves are shown in Fig. 2 when $R_p$=+∞, 200 Ω, 100 Ω, and 50 Ω. All the I-V curves are recorded when the voltage of the source ($V_0$) is swept upwards and downwards, which show the hysteresis behavior when latching happens. The arrows in the figure indicate the sweeping direction of the bias. When $R_p$= +∞ (no $R_p$ in the circuit, i.e. the QCC bias), the normal latching behavior happens with a sharp transition from the superconducting state to the resistive state. The voltage jump is about 187 mV. By decreasing $R_p$, the voltage jump decreases gradually to a few millivolts (3.7 mV at $R_p$=200 Ω), while latching still exists. The slope of the upward transition/latching line in the I-V curve is equal to the internal resistance of the bias to SNSPD and can be expressed as -($R_s$+$R_p$)/($R_s R_p$).

While $R_p$ is less than 100 Ω, a smooth transition is observed instead of the latching behavior, and the hysteresis disappears surprisingly, which is shown as the middle red line of Fig. 2 when $R_p$=50 Ω. Since $R_p$<<$R_s$, it forms a QCV bias to SNSPD. With a further decrease in $R_p$, the I-V curves remain the same with the curve of $R_p$=50 Ω. This result indicates that, in QCV bias, no latching happens to SNSPD. In other words, latching is not an intrinsic characteristic of SNSPD. Instead, it is a result of the improper bias.

Different bias schemes produce various latching behaviors, thus producing different resistive domains after latching. If we compare the resistances of the resistive domains after latching, we notice that the values decrease from 21 KΩ to a minimal value of 255 Ω when the bias changes from QCC bias to QCV bias. Supposing the resistive domains have uniform resistivity and the same width along the nanowire, the QCC and the QCV biases produce resistive domains with lengths of 4.5 μm and 55 nm, respectively. The QCC bias gives the maximal positive feedback energy to SNSPD when latching, while the QCV bias gives the minimal energy that avoids the appearance of latching. Another interesting phenomenon is that the transition points coincide with each other for all the return I-V curves when the bias sweeps back, which is shown as point A in Fig. 2. The resistance of point A is the smallest resistor before SNSPD can jump back to the superconducting state. The downward transition lines, whose slopes are also $-(R_s+R_p)/(R_sR_p)$, are parallel to the respective latching lines. We believe that the value of 255 Ω gives the resistance of the minimal stable resistive domain or hotspot if the nanowire is uniform. The conclusion is that the length of the minimal stable resistive domain is around 55 nm.

Figure 3 shows the nonlatching I-V curve of SNSPD, which can be divided into three regions: the superconducting region, the relaxation oscillation region, and the resistive domain growing region. In the superconducting region, the supercurrent increases with increasing the bias until it reaches the critical current $I_c$. When the current in SNSPD exceeds $I_c$, a gradual and smooth transition from the superconducting state to the resistive domain growing region is clearly seen from the I-V curve. More interestingly, an oscillating signal is observed by using the real-time oscilloscope to monitor the voltage on SNSPD. The origin of the oscillation can be explained to be the dynamic periodic thermal-induced hotspot appearance, growth, and disappearance.[13,14] The inset of Fig. 3 shows a real-time oscillation pulse train with the frequency of around 100 MHz. The oscillation appears only in nonlatching SNSPD with the QCV bias. In the QCC bias, the positive feedback energy is so strong that drives the nanowire from the superconducting state to a steady resistive state directly. No stable oscillation state can exist with QCC bias. A similar oscillation behavior with a much lower frequency in the superconducting microbridge was also reported and analyzed by Skocpol *et al*.[15,16] The *dc* I-V curve shows the averaged values of the oscillation signals. When the bias increases further, the oscillation frequency increases until the thermal balance is reached, which produces a steady resistive domain and terminates the oscillation. Finally, the resistive domain starts to grow and the I-V curve enters the resistive domain growing region.

The single-photon detection performance of SNSPD using the QCV the QCC biases are examined. There are no differences found regarding QE and $R_{dc}$ as well as the jitter. One of the authors previously introduced an anti-latching (AL) method that includes a room-temperature resistor (typically 50 Ω) in parallel with the detector.[17] This method can also avoid the latching behavior of SNSPD. However, it comes with clear disadvantages. On the one hand, the impedance mismatch appears for the response pulse signal, which may produce extra jitter to the output signal. On the other hand, part of the signal is absorbed by the extra resistor, which suppresses the amplitude of the output signal. Figure 4 shows the jitter results measured in three different ways. The jitter for the QCC or QCV bias is 44 ps, which is 20% lower than the jitter of 54 ps measured using the AL method. The inset of Fig. 4 shows the amplitude of the photon response pulse in different cases, which also supports the explanation above.

The foregoing description indicates that SNSPD can run without latching with a QCV bias. The long term running experiment of more than 2 weeks has been done to prove that. Latching never happens no matter how high the bias current is. To further demonstrate the nonlatching nature of SNSPD, we measured the QE of SNSPD with a variable high laser repetition rate. The detector is biased at $I_b=0.95I_c$, where the dark count rate is 10 Hz. A 1550 nm pulsed laser is chosen as the photon source whose repetition rate can be adjusted from 1 MHz to 3 GHz. When the QCC bias is adopted, latching happens when the repetition rate of the photon is 100 MHz or above. SNSPD cannot work even for a second when the repetition rate is over 300 MHz. However, with a QCV bias, SNSPD can work continuously without latching with the pulse frequency up to the laser's upper limit (3 GHz), which is much larger than its maximal counting rate (~ 100 MHz) estimated according to the pulse width. However, the QE is suppressed at high frequencies. This result further proves the nonlatching nature of SNSPD.

In conclusion, latching is not an intrinsic feature of SNSPD but due to the improper bias scheme. By using the QCV bias, the nonlatching operation of SNSPD is realized, and the relaxation oscillation is observed instead of the latching behavior. The SNSPD working in QCV bias shows a better performance than that in the reported anti-latching method with less jitter and higher pulse amplitude. The nonlatching SNSPD assures the unmanned operation of SNSPD in practical applications such as QKD. Besides, it gives us a valuable clue to further understanding the detection mechanism.

1. R. H. Hadfield, J. L. Habif, J. Schlafer, R. E. Schwall, and S. W. Nam: Appl. Phys. Lett. **89** (2006) 241129.
2. L. Ma, S. Nam, H. Xu, B. Baek, T. Chang, O. Slattery, A. Mink, and X. Tang: New J. Phys. **11** (2009) 045020.
3. Y. Liu, T. Y. Chen, J. Wang, W. Q. Cai, X. Wan, L. K. Chen, J. H. Wang, S. B. Liu, H. Liang, L. Yang, C. Z. Peng, K. Chen, Z. B. Chen, and J. W. Pan, Opt. Express **18** (2010) 8587.
4. A. Korneev, A. Lipatov, O. Okunev, G. Chulkova, K. Smirnov, G. Gol'tsman, J. Zhang, W. Slysz, A. Verevkin, and R. Sobolewski: Microelectron. Eng. **69** (2003) 274.
5. K. Suzuki, S. Miki, S. Shiki, Z. Wang, and M. Ohkubo: Appl. Phys. Express **1** (2008) 031702.
6. A. Pearlman, A. Cross, W. Słysz, J. Zhang, A. Verevkin, M. Currie, A. Korneev, P. Kouminov, K. Smirnov, B. Voronov, G. Gol'tsman, and R. Sobolewski: IEEE Trans. Appl. Supercond. **15** (2005) 579.
7. S. Miki, T. Yamashita, M. Fujiwara, M. Sasaki, and Z. Wang: Opt. Lett. **35** (2010) 2133.
8. A. J. Annunziata, O. Quaranta, D. F. Santavicca, A. Casaburi, L. Frunzio, M. Ejrnaes, M. J. Rooks, R. Cristiano, S. Pagano, A. Frydman, and D. E. Prober: J. Appl. Phys. **108** (2010) 084507.
9. J. K. W. Yang, A. J. Kerman, E. A. Dauler, V. Anant, K. M. Rosfjord, and K. K. Berggren: IEEE Trans. Appl. Supercond. **17** (2007) 581.
10. F. Marsili, D. Bitauld, A. Gaggero, S. Jahanmirinejad, R. Leoni, F. Mattioli, and A. Fiore: New J. Phys. **11** (2009) 045022.
11. Y. Korneeva, I. Florya, A. Semenov, A. Korneev, and G. Goltsman: IEEE Trans. Appl. Supercond. **21** (2011) 323.
12. M. Ejrnaes, A. Casaburi, O. Quaranta, S. Marchetti, A. Gaggero, F. Mattioli, R. Leoni, S. Pagano, and R. Cristiano: Supercond. Sci. Technol. **22** (2009) 055006.
13. A. J. Kerman, J. K. W. Yang, R. J. Molnar, E. A. Dauler, and K. K. Berggren: Phys. Rev. B **79** (2009) 100509.
14. R. H. Hadfield, A. J. Miller, S. W. Nam, R. L. Kautz, and R. E. Schwall: Appl. Phys. Lett. **87** (2005) 203505.
15. W. J. Skocpol, M. R. Beasley, and M. Tinkham: J. Appl. Phys. **45** (1974) 4054.

16. A. VI. Gurevich and R. G. Mints: Rev. Mod. Phys. **59** (1987) 941.
17. T. Yamashita, S. Miki, W. Qiu, M. Fujiwara, M. Sasaki, and Z. Wang: Appl. Phys. Express **3** (2010) 102502.

List of figure captions:

Fig. 1. Circuit diagram of SNSPD, which includes a bias circuit and a readout circuit. The resistor parameters are $R_s$=10 KΩ and $Z_0$=50 Ω. $R_p$ can be adjusted from 0 to 1 KΩ. $R_c$=4.8 Ω is the sum of the contact resistance of SNSPD (3 Ω) and the inner resistance of the bias-tee (1.8 Ω).

Fig. 2. I-V curves of SNSPD with different parallel resistors $R_p$=50 Ω, 200 Ω, and 500 Ω, and without $R_p$. The four resistance values indicate the minimal resistance of the resistive SNSPD after switching under different bias schemes. The arrows indicate the sweeping direction of the source. All the data are collected at the temperature of 2.7 K.

Fig. 3. I-V curve of SNSPD with three different regions when $R_p$ is 50 Ω. Three different regions are marked with different background colors. The inset is the voltage pulse recorded from the real-time oscilloscope showing the oscillation with a frequency of 100 MHz in the nanowire at the bias marked with the arrow. The value of the voltage in the inset is amplified by the 50 dB low noise amplifier shown in Fig. 1.

Fig. 4. Timing jitters of SNSPD system with different bias schemes. The jitters are measured under the illumination of the 1550 nm optical light source and the bias current is $0.95I_c$. AL represents the anti-latching method. The inset shows the distinction of the amplitude of the normalized voltage pulses. The dashed lines are guides to the eyes.

Fig. 1

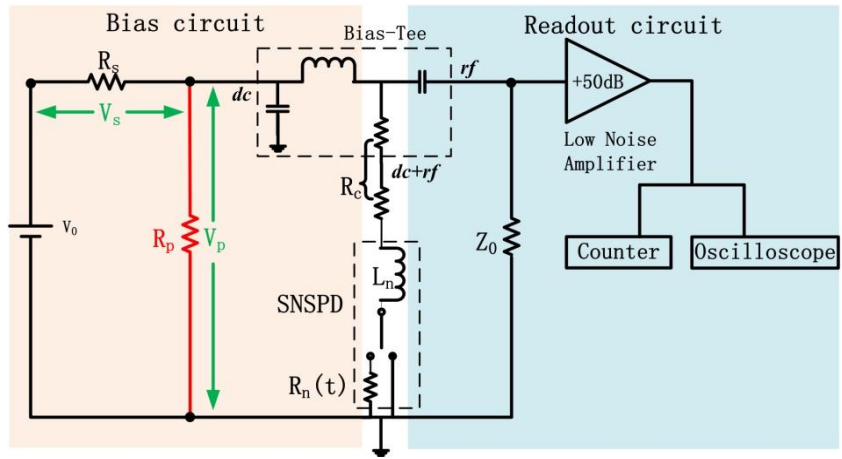

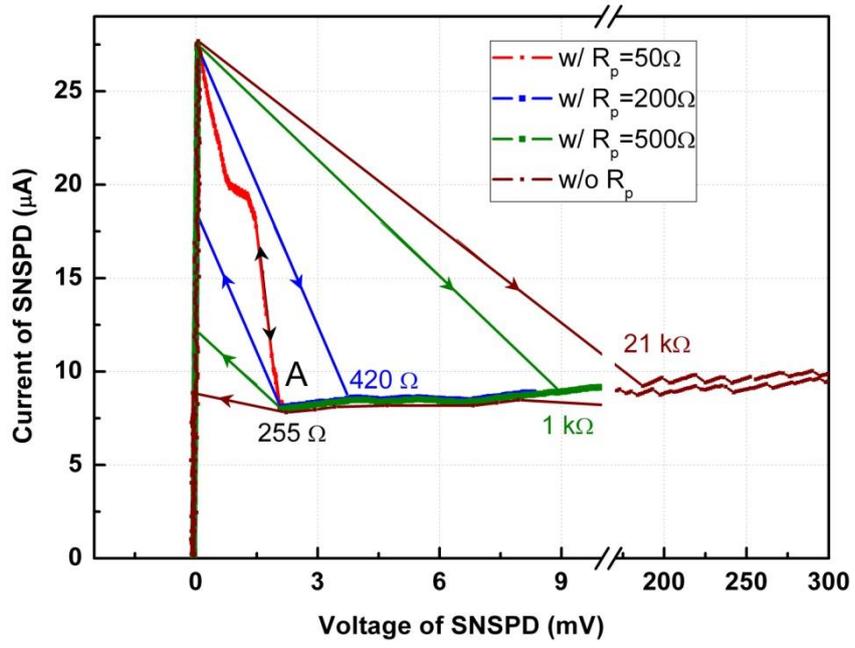

Fig. 2

Fig. 3

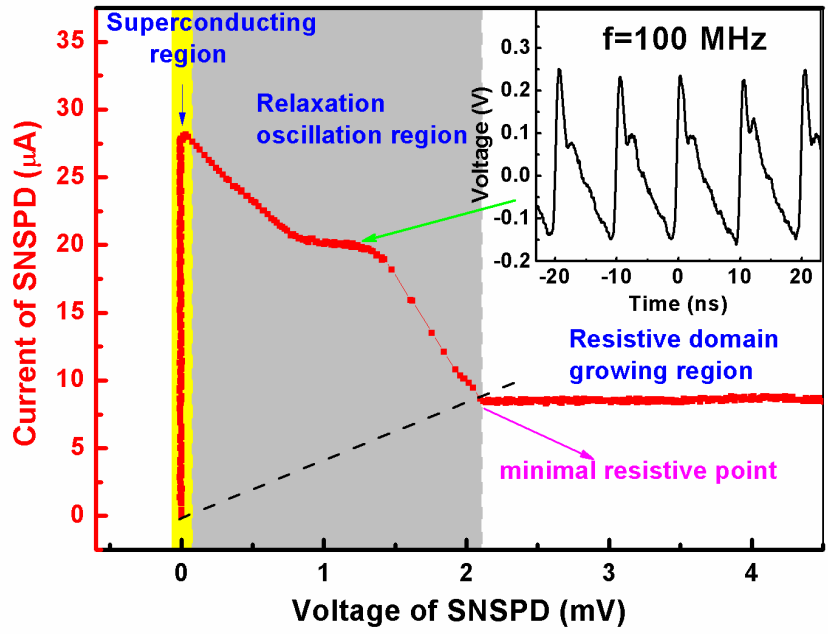

Fig. 4

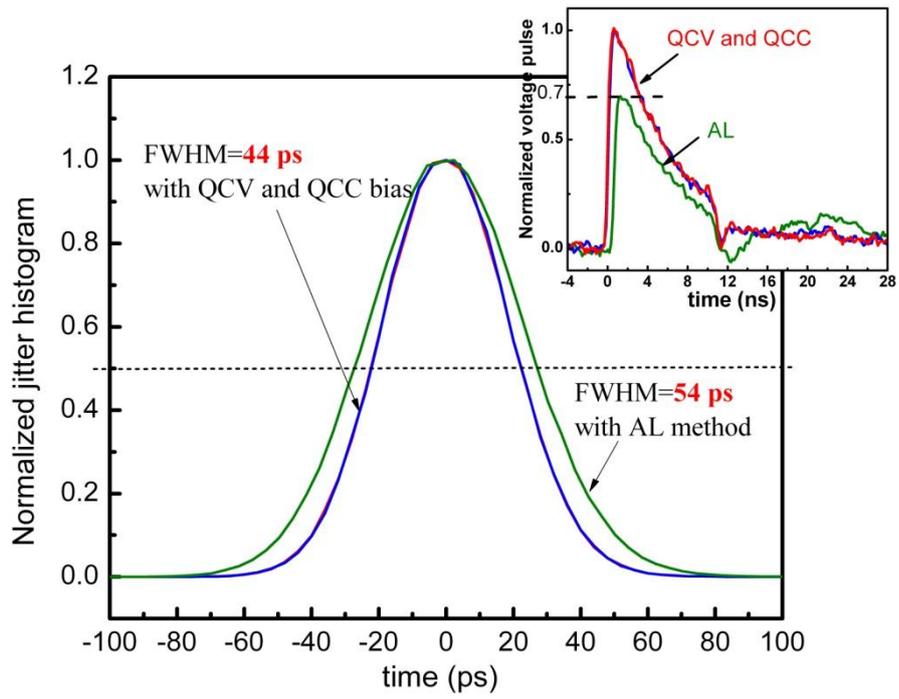